# Potential of Raman scattering in probing magnetic excitations and their coupling to lattice dynamics

**Reshma Kumawat, Shubham Farswan, Simranjeet Kaur, Smriti Bhatia, and Kaushik Sen***

Department of Physics, Indian Institute of Technology Delhi, Hauz Khas, New Delhi 110016, India

*Email: kaushik.sen@physics.iitd.ac.in

**Abstract**

Raman scattering is an excellent method for simultaneously determining the dynamics of lattice, spin, and charge degrees of freedom. Furthermore, polarization selection rules in Raman scattering enable momentum-resolved quasiparticle dynamics. In this review, we highlight the potential of Raman scattering in probing magnetic quasiparticles or excitations in various magnetic materials. We demonstrate how temperature-dependent Raman scattering data can confirm the existence of magnons in long-range ordered magnets and fractionalized excitations in Kitaev spin liquid candidates. To make this review easily understandable to novices, we provide background information on magnons and fractionalized excitations, and explain how they become visible in the Raman scattering process. We also show how to estimate magnetic exchange interactions from the data. For both types of magnetic materials, we discuss the impact of spin-phonon coupling on the lineshape of the phonon modes. In terms of materials, we present magnetic Raman scattering data of antiferromagnetic $Sr_2IrO_4$ and $La_2CuO_4$, ferromagnetic $CrI_3$ monolayers, and Kitaev spin liquid candidates α-$RuCl_3$ and β-$Li_2IrO_3$. Overall, our review demonstrates the versatility of the Raman scattering technique in probing quasiparticles in magnetic quantum materials. The review aims to inform young experimental researchers about the potential of Raman scattering, thereby motivating them to use this technique in their research.

## 1. Introduction

Identifying quantum ground states can be achieved by studying low-energy excitations, particularly through the identification of signature quasiparticles that emerge from these states. Various inelastic scattering methods, such as inelastic neutron scattering, resonant and non-resonant inelastic x-ray scattering, and Raman scattering, have been proven successful in probing these quasiparticles. Among these methods, Raman scattering stands out as exceptional due to its home-laboratory-based setup and the use of a monochromatic laser source, often an affordable visible laser. Moreover, unlike other inelastic scattering methods, especially inelastic neutron scattering, Raman scattering does not require large single crystals. Using a microscope, Raman scattering can probe micron-sized single crystals and thin films down to a few nanometers in thickness [1] [2].

A crucial advantage of Raman scattering is its ability to simultaneously probe scattering responses from lattice, spin, and charge degrees of freedom, thereby revealing fundamental interactions such as electron-phonon and spin-phonon interactions. Additionally, the dynamical response in Raman scattering lies in the THz domain, which is crucial for probing low-energy excitations close to the Fermi energy. Furthermore, the Raman scattering response strongly depends on the polarization of incident and scattered photons and the underlying lattice geometry, allowing the dynamics of quasiparticles in crystals to be probed as a function of their momentum.

In this review, we focus on the magnetic excitations in long-range ordered magnets and in quantum spin liquid (QSL) systems without long-range order but with strong quantum fluctuations. We demonstrate with examples that Raman scattering can efficiently probe two-magnon excitations in antiferromagnets, one-magnon excitations in ferromagnets, and fractionalized excitations, such as Majorana fermions in Kitaev candidates for QSL. The success of this review lies in providing information to researchers working on magnetic quantum materials about the potential of polarization-resolved low-temperature Raman scattering. Our target audience includes young experimental scientists who are currently working, or aspire to work, in low-temperature polarized Raman scattering to explore magnetization dynamics in quantum materials.

We intentionally avoid delving into the complex theoretical background of Raman scattering processes. Instead, we aim to provide straightforward explanations to help researchers understand and utilize Raman scattering for investigating the magnetic ground states of quantum materials. For those interested in a more detailed theoretical understanding, we recommend several exceptional reviews that capture significant milestones in the field of Raman scattering studies of quantum materials. Notably, T. Devereaux and R. Hackl wrote a comprehensive review on how Raman scattering can be used to probe momentum-resolved charge dynamics in correlated electronic systems [3]. P. Lemmens, G. Güntherodt, and C. Gros reviewed Raman scattering from magnetic excitations in low-dimensional spin systems [4]. D. Wulferding et al. provided a perspective on Raman scattering studies of fractional excitations in QSL systems [5].

The structure of our review is as follows: Section 2 describes magnons in long-range ordered magnets and fractionalized excitations in 2D Kitaev QSL systems, highlighting the essence of spin-phonon coupling. Section 3 describes a basic Raman scattering setup to facilitate understanding of the Raman scattering data discussed in the following sections. Section 4 summarizes the challenges in performing Raman scattering experiments and suggests possible methods to address these challenges for accurate interpretation of the Raman scattering data. Section 5 focuses on one-magnon and two-magnon Raman scattering mechanisms. Section 6 presents selected experimental Raman scattering data for magnetic materials, showcasing how Raman scattering can effectively study magnetic quasiparticles. Finally, Section 7 provides a summary and a perspective.

## 2. Magnetic excitations and their coupling to lattice degrees of freedom

We discuss two types of magnetic excitations: magnons (bosons) in long-range ordered ferro and antiferromagnets, and fractionalized excitations (fermions) in Kitaev systems with enhanced spin fluctuations in absence of long-range orders.

### 2.1. Magnons

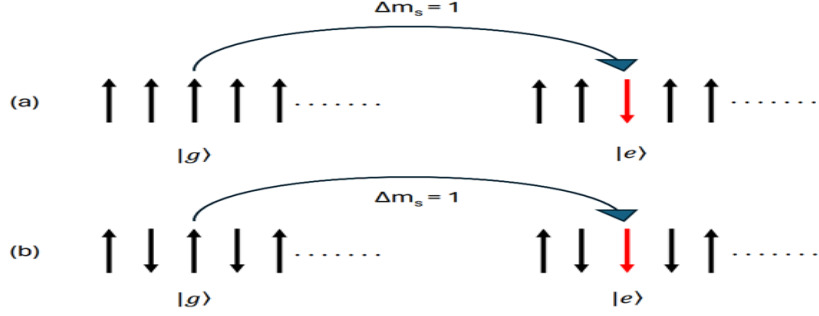

**Figure 1.** (a) Left: Ising-type ferromagnetic ground state ($|g\rangle$) with all the spins in up-direction. Right: The first excited state ($|e\rangle$) for the corresponding ferromagnet. The total change of spin from $|g\rangle$ to $|e\rangle$ is $\Delta m_s = 1$. (b) The corresponding figure for antiferromagnetic spin chain.

Crystalline magnetic materials achieve long-range magnetic orders due to exchange interaction. However, perfect magnetic orders are obtained only at $T = 0$ K. Any finite temperatures disturb the ground state by generating elementary excitations of integer spins in ferro and antiferromagnets. The corresponding quasiparticles are known as magnons that follow Bose statistics. Figure 1(a) represents a one-dimensional (1D) ferromagnetic ground state ($|g\rangle$) with all spins in the same direction, i.e., up in the present case. The first excited state ($|e\rangle$) involves a flipped electronic spin in the down direction. The same applies to a 1D antiferromagnet, as shown in Figure 1(b). Obviously, flipping a spin costs energy which can be determined from the following Heisenberg Hamiltonian with the Ising spins.

$$H = -2J \sum_i \vec{S}_i \cdot \vec{S}_{i+1}$$ 

Eq. (1)

Here, $J$ is the nearest neighbour exchange interaction ($J > 0$ for a ferromagnet and $J < 0$ for an antiferromagnet) and $\vec{S}_i$ is the spin state at $i^{\text{th}}$ site. For simplicity, we neglected exchange interactions beyond the nearest neighbor. Considering, $N$ number of spins of magnitude $S$ in the 1D linear chain, we obtain the ground state energy, $E_0 = -2J \sum_{i=1}^{N} \vec{S}_i \cdot \vec{S}_{i+1} = -2NJS^2$ for a ferromagnet. In the first excited state, the energy is $E_1 = -2J \sum_{i=1}^{N-2} \vec{S}_i \cdot \vec{S}_{i+1} + 2 \times 2JS^2 = -2(N-2)JS^2 + 2 \times 2JS^2 = -2NJS^2 + 8JS^2$. The energy difference between the ground and excited states for the ferromagnetic system is $\Delta E = E_1 - E_0 = 8JS^2$. One can get the same energy difference for the antiferromagnetic case, i.e. $-8JS^2$ (it is positive as $J < 0$ for antiferromagnets). On the other hand, the overall change in the total spin is $\Delta m_s = 1$ which is an integer number. Thus, the involved quasiparticles with these excited states, known as 'magnons', have the energy of $8|J|S^2$ and the integer spin 1 (bosons).

Magnons are quasiparticles that arise from the quantization of spin waves in crystal lattices. Here, using a semiclassical approach, we will explain the occurrence of a spin wave in the 1D ferromagnetic lattice. As shown in Figure 1(a), the ferromagnetic order is characterized by all the spins in the same direction, i.e. 'up' in the present case. Here, the spins are aligned in the same direction in response to the internal magnetic field generated by exchange interactions among the spins. Certainly, such a perfect alignment could only be achieved at $T = 0$ K. Any finite temperatures cause deviations from the perfect alignment, which paves the way for the internal magnetic field to yield a torque on to the spins.

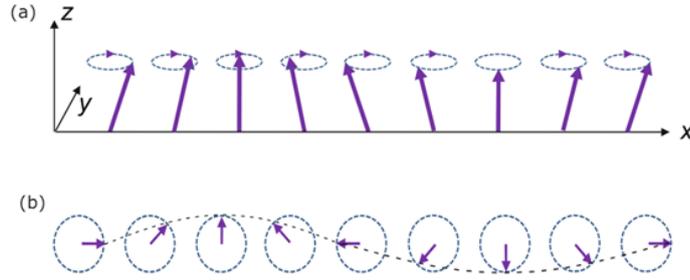

**Figure 2.** (a) Precession motion of the spins around the quantization axis $z$. The 1D ferromagnetic spin chain is along $x$-axis. (b) The pattern of the spin components on the $xy$-plane. The dotted line connecting the spin tips represents propagating spin wave along $x$-axis. The picture is taken from [34].

Thus, the spins develop a precession motion around the quantization axis, i.e. the direction of the internal field, as depicted in Figure 2 (a). In the figure, $z$-axis is the quantization axis. If we draw arrows from the $z$-axis to the tip of the spins, the arrows on the $xy$-plane manifests a counterclockwise rotation, as show in Figure 2(b). The dotted line connecting the spin tips represent a periodic variation of the $y$-component of the spin moments along the $x$-axis. This is the propagating spin wave along the $x$-axis that effectively arises from the collective motion of all the spins in a periodic lattice. The dispersion relation, i.e. energy ($\hbar\omega$) versus wavevectors ($q$) of this spin wave is [6]:

$$\hbar\omega = 4JS(1 - \cos qa).$$  Eq. (2)

Here, $a$ is the lattice parameter in the 1D lattice and $q$ is the wavevector. Figure 3(a) shows the dispersion relation from the Brillouin zone center to one of the Brillouin zone boundaries. The symmetric relationship applies for $q < 0$. Magnon energy decreases to 0 at long wavelength limits. At $q \to 0$, $\hbar\omega$ goes to 0, indicating no magnon energy gap at the Brillouin zone centre. Whereas at $q \to \pi/a$, the magnon energy tends to $8JS^2$ along the line of the energy difference between the ground and the first excited states of a ferromagnetic spin chain.

So far, our discussion assumes isotropic exchange interaction $J$ and the absence of any external magnetic field and internal magneto-crystalline anisotropy. If we consider both, the effective Hamiltonian can be written as [7].

$$H = -\sum_{ij} J_{ij}(\vec{S}_i \cdot \vec{S}_j) + \gamma\hbar\sum_i \vec{S}_i \cdot \vec{B} - K\sum_i (\vec{S}_i^z)^2.$$  Eq. (3)

Here, $\gamma = g\mu_B/\hbar$ is the gyromagnetic ratio, $g$ is the Lande factor, $\mu_B$ is the Bohr magneton, $\hbar$ is the reduced Planck constant, $\vec{S}_i$ is the spin at the lattice site $i$, $\vec{B}$ is the external magnetic field, and $J_{ij}$ is the exchange interaction between spins $\vec{S}_i$ and $\vec{S}_j$. $K$ is magneto-crystalline anisotropy energy along (single-ion anisotropy) the $z$-axis.

Using linear spin wave theory [8] based on the above Hamiltonian, we calculated magnon dispersion for a ferromagnetic spin chain with $s = 1/2$. We assume only nearest neighbor exchange interaction of $J = 1$ meV. Further, direction of the applied magnetic field and the relevant magneto-crystalline anisotropy are assumed to be perpendicular to the spin chain. In presence of the last two terms in Eq. (3), a magnon gap opens at the zone-center, as shown in Figure 3. This magnon gap increases with the magnitude of the applied magnetic field and with the underlying magneto-crystalline anisotropy (also known as single-ion anisotropy that gives rise to a preferential orientation of the spins). The effect of magneto-crystalline anisotropy on magnon dispersion is much stronger than the applied magnetic field.

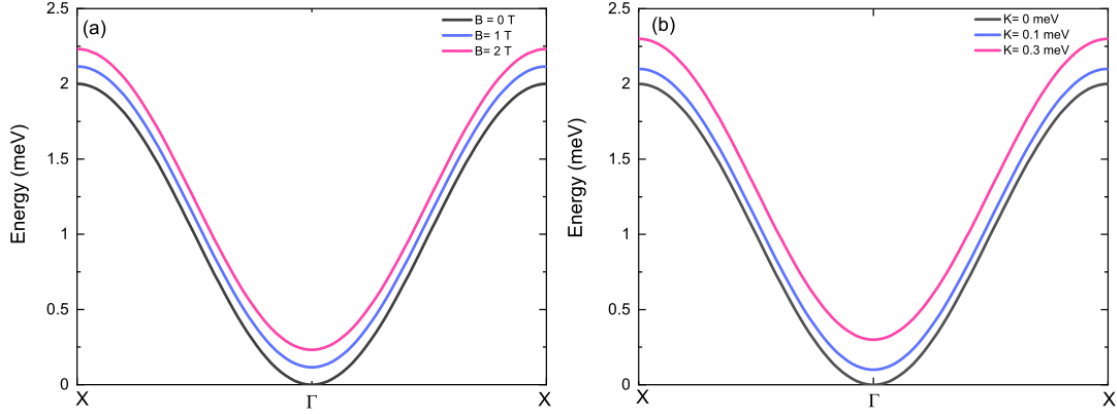

**Figure 3.** (a) Magnon dispersion for a ferromagnetic spin chain by considering only the exchange interaction of $J = 1$ meV between the neighboring spins ($s = 1/2$) at several magnetic fields ($B$). (b) The corresponding magnon dispersion for several single-ion anisotropy constants ($K$) at zero magnetic field. A gap opens in the magnon dispersion at the zone-center due to finite $B$ or $K$.

## 2.2. Fractionalized excitations

So far, we talked about magnetic systems with long-range ferro and antiferromagnetic orders that have the elementary excitations of magnons which follow Bose-Einstein statistics. In this section, we discuss magnetic materials where the presence of strong quantum fluctuations prevents long-range ordering. The origin of these strong quantum fluctuations is specific to the underlying lattice geometry. One famous example is the triangular antiferromagnets with Ising spins at the vertices of the triangles. On such a triangular motif, we cannot simultaneously satisfy the antiparallel spin alignment of the consecutive spins. Hence, the triangular antiferromagnetic network remains frustrated and develops strong quantum fluctuations that prohibit any long-range orders. Such a ground state is known as quantum spin liquid [9].

Of particular interest is the Kitaev toy model for $S = 1/2$ spins arranged on two-dimensional (2D) honeycomb lattice with nearest-neighbor Ising interactions that have the bond dependent magnetic easy axes, as shown in Figure 4. Due to three different anisotropic Ising interactions, a spin, for instance at the position #1 in Figure 4, can be in three different degenerate states, i.e., the three different alignments of the spin are equally possible. Since all the magnetic sites are equivalent, the same scenario holds for all other spin sites. As a result of this magnetic frustration, no specific order can be found down to very low temperatures, theoretically down to 0 K [10]. The characteristic feature of such a QSL ground state is the elementary fractionalized spin excitations. In Kitaev model, the elementary spin $S = 1/2$ fractionalizes into two classes of Majorana fermions: itinerant Majorana fermions and localized $Z_2$ fluxes. Inelastic light scattering has been proven as an efficient method to probe the itinerant Majorana fermions, while the detection of $Z_2$ fluxes is experimentally challenging [5]. The itinerant Majorana fermions contribute to inelastic light scattering process in two ways. First, the creation or annihilation of a pair of fermions. Second, creation of one fermion and the annihilation of the other fermion. The temperature-dependence of the scattering response in these two processes are distinct. The spectral weight for the first one is proportional to $[(1 - f(E_1))(1 - f(E_2))\delta(\hbar\omega - E_1 - E_2)]$, and for the second one, it is $[f(E_1)(1 - f(E_2))\delta(\hbar\omega + E_1 - E_2)]$ [11]. Here, $f(E)$ is the Fermi-Dirac distribution function, $\hbar\omega$ is the Raman shift, $E_1$ and $E_2$ are the energies of the two fermions.

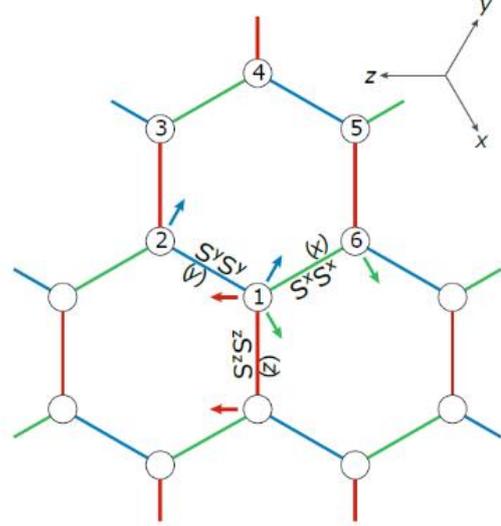

**Figure 4.** Kitaev spin model for s=1/2 spins on 2D honeycomb lattice. The green bonds have easy axis along $x$-axis. The blue bonds have easy axis along $y$-axis. The red bonds have easy axis along $z$-axis. The figure is adapted from ref. [10].

## 2.3. Spin-phonon coupling

Elastic deformations in magnetic materials can affect magnetization of the systems, given that the spins and lattice are strongly entangled, e.g. via the spin-orbit interaction. Such systems show macroscopic magnetostriction or magnetoelastic effects. Notably these are static properties of the systems. Here, however, we talk about the coupling between the dynamical quasiparticles phonons and spin excitations (magnons and fractionalized excitations). Spin excitations contribute to phonon self-energy, $\Sigma = \Sigma' + i\Sigma''$. The real part of the self-energy, $\Sigma'$ determines renormalized phonon frequency, and the imaginary part $\Sigma''$ determines the phonon lifetime which is the inverse of phonon linewidth. In general, when optical phonon modes are coupled to the underlying spin excitations in magnetic system, the spin-phonon Hamiltonian can be written as[12]:

$$H = H_{\text{spin}} + H_{\text{spin-phonon}} + H_{\text{phonon}}, \qquad \text{Eq. (4)}$$

where $H_{\text{spin}}$ is the bare spin Hamiltonian of the system, $H_{\text{phonon}}$ is the bare lattice dynamics part, and $H_{\text{spin-phonon}}$ is the coupled Hamiltonian when spin-phonon coupling is finite.

Lattice vibrations, i.e., phonons lead to the dynamic changes in the underlying exchange interactions $J$ that can be expressed as per the following.

$$J_{ij} = J_0 + \frac{\partial J_{ij}}{\partial u_i} u_{ij} + \frac{1}{2} \frac{\partial^2 J_{ij}}{\partial u_{ij}^2} u_{ij}^2 \qquad \text{Eq. (5)}$$

Here, $J_0$ is the equilibrium exchange coupling constant, $\vec{u}_i$ is the displacement of the $i^{\text{th}}$ atom, $\vec{u}_{ij} = \vec{u}_i - \vec{u}_j$. The first term in Eq. (5) gives rise to bare spin Hamiltonian, $H_{\text{spin}} = \sum_{ij} J_0 \vec{S}_i . \vec{S}_j$. The last two terms constitute $H_{\text{spin-phonon}}$.

Renormalization of phonon frequency is proportional to average spin-spin correlation. However, in solids, not all phonon modes are renormalized due to underlying magnetism. If the atomic displacements associated with a phonon mode modulate the underlying magnetic exchange interactions, then the phonon mode is said to have a significant spin-phonon coupling. The change in phonon frequency due to renormalization can be expressed in terms of the following expression [13], [14].

$$\Delta\omega = -\frac{1}{2\mu_\alpha \omega_\alpha} \sum_i \frac{\partial^2 J_{i,i+1}}{\partial u_\alpha^2} \langle \vec{S}_i . \vec{S}_{i+1} \rangle \qquad \text{Eq. (6)}$$

Here, $\omega_\alpha$ is the mode frequency, $\mu_\alpha$ is the reduced mass associated with the mode, $J_{i,i+1}$ is the nearest neighbor exchange interaction, $u_\alpha$ is the atomic displacements associated with the phonon mode, and $\langle \vec{S}_i . \vec{S}_{i+1} \rangle$ is the spin-spin correlation function. In the presence of spin-phonon coupling, phonons not only decay via interactions with other phonons, but also, they interact with magnons. Thus, the phonon lifetime will be shorter in magnetic systems, which leads to broader linewidth of the phonons with finite spin-phonon coupling.

## 3. Basic Raman scattering setup and an exemplary Raman spectrum

Figure 5 illustrates a schematic of a Raman scattering setup that records Raman scattering responses in quasi-backscattering geometry. To familiarize the authors with the experiment, essential components are given here. The Raman scattering is a photon-in and photon-out process. The first step involves exciting samples with a monochromatic excitation laser, typically in the visible range, especially for the study of quantum materials as discussed in this review paper. The final step includes the recording of the scattered light of different wavelengths other than original laser excitation wavelength. Thus, a Raman scattering setup is designed to record inelastic light scattering. Additionally, by controlling the polarization of incoming and outgoing photons, we can probe scattering response of defined symmetry for a crystallographic point group.

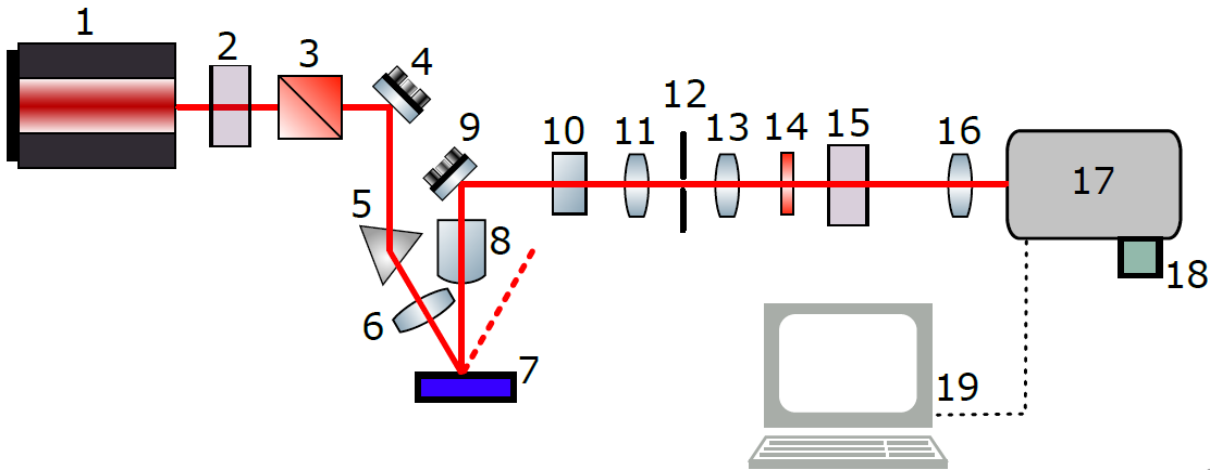

**Figure 5.** Basic Raman scattering setup in a quasi-backscattering geometry. All components are described in text.

Component-1 (C-1) is the monochromatic excitation laser with narrow bandwidth. C-2 is a linear polarizer. After C-2, the linear polarized light passes through a Soleil-Babinet compensator (C-3), followed by reflection off a mirror (C-4) and passing through a right-angle prism (C-5). This arrangement ensures that the incident light is focused through an achromat (C-6) onto the samples (C-7) at an angle of approximately 30°. The scattering response is collected with a long working distance objective (C-8) along the surface normal of the sample. Since the incident light and the outgoing light do not coincide, i.e. they do not have a perfect 180° angle between them, this setup is said to have quasi-backscattering geometry. Most commercial Raman spectrometers, however, use a perfect backscattering geometry.

To ensure that the incoming photon polarization is parallel to the sample surface, a Soleil-Babinet compensator is employed to rotate the polarization axis. This is important as the incident light does not hit the sample surface along the surface normal. This quasi-backscattering geometry is effective to strongly reduce the elastic scattering response. The collected Raman response by the objective (C-8) is guided via a mirror (C-9). Subsequently, a Bragg filter (C-10) is introduced in the scattered beam path to further diminish the elastic Raman response. Subsequently, the scattered light passes through a confocal aperture accompanied by two achromats (C-11, 12, 13). Such a confocal geometry is used to reduce scattering background by confining the probe depth near the sample surface.

The combination of a quarter waveplate (C-14) and an analyzer (C-15) selects the polarization of the scattered light. Finally, scattered light with selected polarization enters a grating spectrometer (C-17) through an achromat (C-16). Raman scattering responses are recorded using a two-dimensional CCD detector (C-18). Finally, the PC (C-19) connected to the spectrometer saves the Raman scattering intensity data as a function of Raman shift or wavelength. The Raman shift is defined as $\hbar(\omega_i - \omega_s)$, where $\omega_i$ and $\omega_s$ are the frequencies of the incident and scattered photons, respectively.

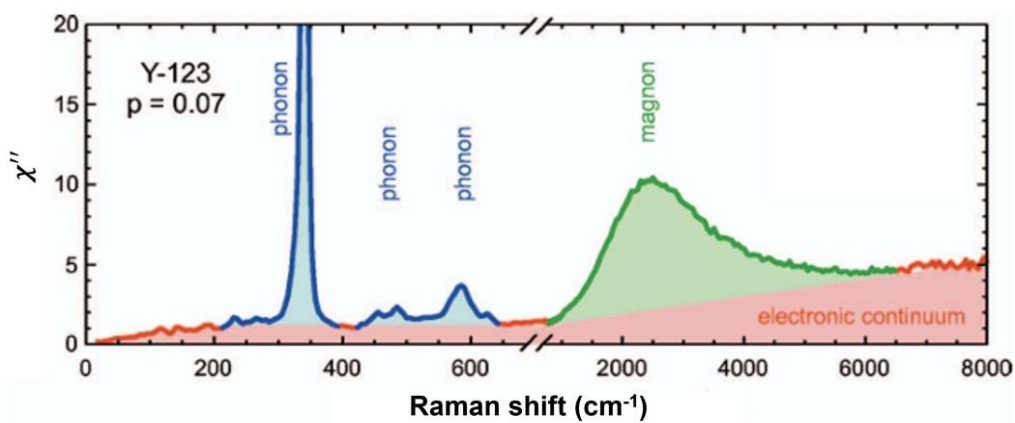

**Figure 6.** The Raman scattering response from an underdoped cuprate compound over a wide energy-range. The spectrum comprises of sharp phonon peak, a relatively broad two-magnon mode, and the background scattering continuum arises from conduction carriers. The figure is adapted from [3].

Figure 6 shows Raman scattering response ($\chi''$) from an underdoped cuprate, precisely $Y_{0.92}Ca_{0.08}Ba_2Cu_3O_{6.3}$ which has 0.07 charge carriers per $CuO_2$ planes [3]. Such a single Raman spectrum simultaneously probes three different dynamics in the system: lattice dynamics (phonons), spin dynamics (magnons), and charge dynamics (electron-hole pairs). The sharp modes colored in blue are the phonons arising from the lattice dynamics. Relatively at high energy, the broad mode shaded in green is two-magnon Raman scattering that arises from the spin dynamics in the antiferromagnetic ground state. The Raman spectral background in red comes from the dynamics of mobile charges. In the case of insulators, such a Raman spectral background would be zero. As will be discussed later, the Raman spectral background may arise from spin excitations in the magnetic insulating compounds. However, the temperature-dependence of such spectral background would be very different from the one for mobile charge carriers.

## 4. Experimental challenges and solutions in Raman scattering experiments

It is a known fact that acquiring Raman scattering data from a system is relatively simple, as described in the setup outlined in Section 3. However, the main challenge lies in the interpretation of the Raman scattering response. The accuracy of this interpretation depends on how precisely we determine the pure Raman scattering response of a system, taking into account all experimental parameters that may adversely influence the scattering response. In this section, we discuss the prevalent challenges in determining the Raman scattering response and methods to mitigate these challenges.

The signatures of electronic and magnetic Raman scattering are embedded in the Raman spectral background. For instance, mobile charge carriers in metallic systems yield a continuum Raman scattering background that is weakly energy-dependent, fractionalized excitations in quantum spin liquid systems also yield a similar background that follows Fermi-Dirac statistics, and two-magnon Raman scattering from antiferromagnets often gives rise to a broad Raman mode that can be misinterpreted as a spurious Raman spectral background arising from the instrument, sample environment, or fluorescence signal from the sample. Hence, the first task is to correctly determine the finite Raman spectral background of a Raman scattering setup that is temperature- and sample-independent. To do this, we need to measure a non-magnetic insulator that has a finite number of phonon modes. We do not expect any continuum Raman scattering (Raman spectral background) from such a sample. Therefore, the measurement of such a sample should yield Raman phonon modes with zero spectral background. This allows us to determine the spurious Raman spectral background from the instruments by subtracting the observed phonon modes. The instrumental background, including the background contribution from the sample environment, can be polarization-dependent in some cases. Thus, for polarization-dependent Raman scattering experiments, we need to measure this insulating sample in the required polarization configurations and determine the Raman spectral background in all these polarization geometries. Additionally, excitation lasers can yield sample-dependent fluorescence backgrounds that complicate the expected Raman scattering continua. When analyzing electronic Raman scattering data and the scattering response from quantum spin liquid systems, it is necessary to measure the sample with a second excitation laser of a different wavelength. If the second excitation laser also yields the same Raman scattering continuum background, then we can trust such a spectral background.

Another experimental challenge related to Raman continuum scattering is the energy dependence of the Raman spectral background. In any Raman scattering setup, the Raman scattering response is a function of the Raman shift over a wide range of energy scales. For example, in Figure 5, this response depends on the CCD detector, as the detection efficiency is strongly wavelength-dependent. We can accurately determine such energy dependence with the help of a well-calibrated blackbody radiation source. When we record a spectrum from such a white light source, the output signal will be convoluted with the wavelength-dependent efficiency of the CCD. By dividing the output signal by the input white light signal, we determine the wavelength-dependent normalization factor of the instrument. For any subsequent Raman spectra, we can use this normalization factor to correct the energy-dependent spectral background.

Laser-induced heating effect is another concern in Raman scattering experiments that prevents us from measuring samples down to the base temperatures of cryostats. This heating effect is unavoidable; however, it can be minimized by using low laser powers. In microscopic Raman setups (commercial Raman spectrometers), we typically measure with less than 1 mW laser power, for which the focused laser spot size on the sample surface is 5-10 μm in diameter. In contrast, for macro-Raman setups (i.e., without a microscope objective), the incident laser power can be significantly higher as the laser spot size is 60-100 μm in diameter. In addition to keeping the laser power low, we need to carefully determine the resultant temperature within the laser spot area on the sample during Raman scattering experiments. This is particularly crucial for the analysis of the Raman scattering response from mobile charges and fractionalized excitations. To determine effective sample temperature during measurements, we need to use a low-energy Raman phonon mode. Specifically, we need to record both the Stokes and anti-Stokes lines of the corresponding mode. The intensity ratio of these two lines follows Maxwell-Boltzmann statistics. The equation for the intensity ratio of Stokes ($I_s$) to anti-Stokes ($I_{as}$) lines is given by: $I_{as}/I_s = \exp(-\hbar\omega/k_B T)$, where $\hbar\omega$ is the energy of the phonon mode, $k_B$ is the Boltzmann constant, and $T$ is the effective temperature. Note that, we should determine the effective temperature following this method at all desired temperatures of our measurements.

Finally, we would like to mention that, although the Raman scattering method is powerful, meaningful data analysis often requires complementary theoretical calculations and experiments. For example, for the analysis of Raman phonon modes, in addition to polarized Raman measurements, first-principles calculations of lattice dynamics lead to accurate mode assignment. To extract magnetic exchange interactions from two-magnon and one-magnon Raman scattering, simulations based on linear spin wave theory are essential. In addition to calculations, inelastic neutron scattering and resonant inelastic x-ray scattering data help to analyze and interpret Raman scattering data, as those experiments obtain spin dispersion relations. Furthermore, these experiments can also identify signatures of fractionalized excitations from quantum spin liquid candidates [15].

## 5. One-magnon and two-magnon Raman scattering

In this section, we discuss the underlying mechanisms of one-magnon and two-magnon Raman scattering in the context of long-range ordered ferromagnets and antiferromagnets, respectively. We would like to highlight a few points here. A sizable Raman scattering response can arise from magnetic excitations in a ferromagnet via one-magnon Raman scattering, provided the magnetic ions possess a significant spin-orbit interaction. In comparison to one-magnon Raman scattering, two-magnon Raman scattering is much stronger in antiferromagnets, which maps out the magnon density of states in the antiferromagnetic phase. As will be explained later, two-magnon Raman scattering cannot occur in a ferromagnet.

## 5.1. One-magnon Raman scattering

One-magnon Raman scattering involves two photons and one magnon. In Stokes process, an incoming photon (frequency $\omega_i$, momentum $\vec{k}_i$) interact with an electron spin and flip the spin that yields a single magnon ($\omega_m$, $\vec{q}$). In the process, a scattered photon ($\omega_s$, $\vec{k}_s$) is generated with reduced energy in comparison to the energy scale of the incident photon. The conservation of energy and momentum yields: $\hbar\omega_i = \hbar\omega_m + \hbar\omega_s$ and $\hbar\vec{k}_i = \hbar\vec{q} + \hbar\vec{k}_s$. The momentum conservation dictates which magnon in the Brillouin zone will be created. Typical Raman experiments we perform in visible range. Therefore, momentum given to the electronic spin by the incident photon of energy 532 nm, for instance, is $\sim 2\pi/\lambda \approx 1.2 \times 10^{-3}$ Å$^{-1}$. The momentum of magnon will be $q \sim 10^{-3}$ Å$^{-1}$. If we consider lattice spacing (distance between consecutive spin sites) in the one-dimensional spin chain is 1 Å, then the Brillouin zone boundary appears at 3.14 Å$^{-1}$. Thus, we will be able to generate a magnon or a magnetic excitation only at the Brillouin zone centre, provide there is a gap in the spin dispersion at the zone centre, as shown in Figures 3(a) and (b). In what follows is how this one-magnon scattering process takes place.

Light can couple to electronic spins in two ways. The direct coupling is the magnetic dipole interaction of the magnetic field of light to the electronic spin of the magnetic ion. The second one is the indirect coupling that involves electric dipole interaction of the electric field of the light to the electron of the magnetic ions. Both experiments and theory have shown that the later one is much stronger than the direct coupling. This indirect coupling occurs through the mixing of the spin and orbital motion of electrons, i.e., via spin-orbit coupling [16][17]. We will understand the one-magnon Stokes-part of the one-magnon Raman scattering process through this following example shown in Figure 7. Let's consider a ferromagnetic ground state which has the total spin momentum of $S$ and the total orbital momentum of $L = 0$. However, the immediate excited state of the system has the orbital angular momentum of $L = 1$ and the same spin angular momentum of $S$. Here, we consider that the excited has the strong spin-orbital interaction ($\lambda$). Thereby, the excited state is splitted into three $J$ states, i.e., $J = S - 1, S, and\ S + 1$. Although, we don't consider $\lambda$ in the ground state, the exchange interaction among the spins, or the applied magnetic field lift the degeneracy of the ground state, and lead to various non-degenerate state having the total spin momentum from $S$ to $-S$ in steps of $-1$. The energy difference between the low-lying states $S$ and $S - 1$ is equal to the energy of one magnon $\hbar\omega_m$. In comparison, the energy of the incident visible laser is much higher than the magnon energy.

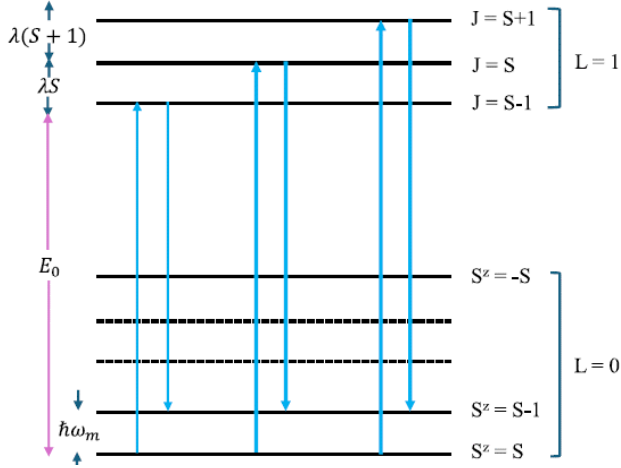

**Figure 7.** Principle of one-magnon Raman scattering for a ferromagnet with strong spin-orbit interaction ($\lambda$). $E_0$ is the energy difference between the ground and the first excited state. One magnon energy is equal to $\hbar\omega_m$. The figure is adapted from ref. [18].

As illustrated in the Figure 7, one-magnon Raman scattering is associated with the electronic transition from the lowest $S$ state to the next $S - 1$ state. Here, one can show using third-order perturbation theory that the excited spin-orbit mixed states work as the intermediate states for the transitions. Thus, for practical Raman scattering experiments, we expect to see one-magnon Raman scattering when the magnetic ions possess sufficient spin-orbit interaction. For clarity, we want to stress that one-magnon Raman scattering can be seen in both ferromagnets and antiferromagnets. However, as will be explained in the following, the two-magnon Raman scattering is the dominant scattering mechanism in antiferromagnets, and such scattering does not happen in a linear ferromagnet.

## 5.2. Two-magnon Raman scattering

As the name suggests, two-magnon Raman scattering involves the creation and annihilation of two magnons in Stokes and anti-Stokes Raman scattering processes, respectively. Let's consider the case for Stokes Raman scattering. It involves an incident photon ($\hbar\omega_i$), a scattered photon ($\hbar\omega_s$), and a pair magnons. The energy and momentum conservation suggests: $\hbar\omega_i = \hbar\omega_s + 2\hbar\omega_m$ and $\hbar(\vec{k}_i - \vec{k}_s) = \hbar(\vec{q}_1 + \vec{q}_2)$. As explained before, $k = |\vec{k}_i - \vec{k}_s|$ is negligible in comparison to the Brillouin zone boundary. Hence, $\vec{q}_1 \approx -\vec{q}_2$ which indicate the creation of two magnons with opposite momentums. From the spin wave dispersions given in Figure 3, magnons of opposite momentums will have the same energy. Figure 3 further suggests that there is no restriction on which two magnons participate in the scattering process. If their momentum is opposite, they take part in this scattering process. Hence, two-magnon Raman scattering obtains the information of the entire magnon density of states. However, as displayed in Figure 3, spin wave dispersion is flatter towards the zone-boundary which leads to the large density of states of magnons. Thus, zone-boundary magnons contribute maximum to the two-magnon scattering. In summary, energy-dependence of two-magnon Raman scattering mode is proportional to the magnon density of states. However, it is not a strict proportionality. Instead, a polarization-dependent (polarization of incoming and scattered photons) and wavevector-dependent weighting function is involved due to the nature of coupling of the magnons with the incident light. This is explained in detail in ref. [18]. Inelastic neutron scattering often complements magnetic Raman scattering data, as the technique probes spin dispersion relations (magnon energy as a function of crystal momentum). Raman scattering, however, has much higher energy resolution (~0.06 meV) with respect to inelastic neutron scattering (~1-3 meV). The advantage of Raman scattering is that it can measure micron-size samples, whereas inelastic neutron scattering requires centimeter-size samples.

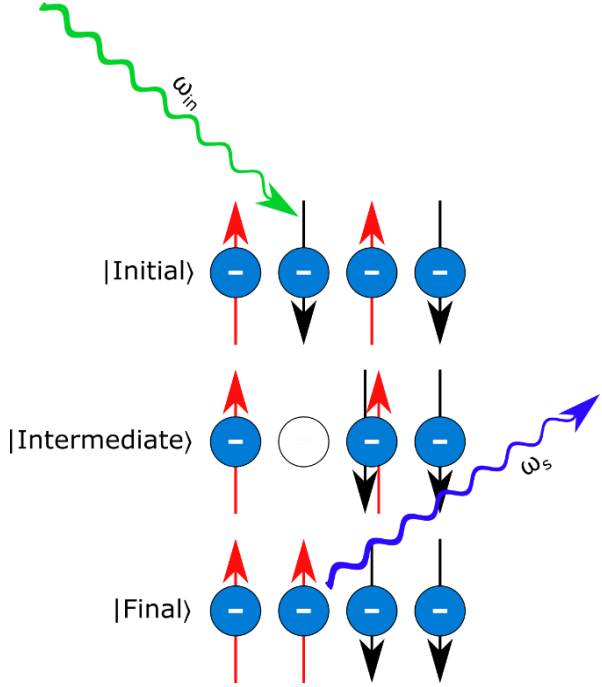

The most relevant mechanism for two-magnon Raman scattering is the excited-state exchange interaction. To understand this, we will simply consider an antiferromagnetic spin chain. The antiferromagnetic spin chain comprises of two magnetic sublattices of spin up and down. They are shifted by one-lattice spacing. The electric field of light gets coupled with an electronic spin via electric-dipole interaction. In an intermediate state, the electron hopped to a neighboring atomic site, leading to a double occupancy. Since this is not a favorable state due to onsite Coulomb repulsion, one of the possibilities is that the electron in the second atomic site comes to occupy the vacant position in the former atomic site. Overall, the total spin change between the initial and final state is $\Delta S = 0$. However, locally there are two spin flips. In this system, the energy-scale of the two-magnon Raman scattering would be the energy difference between the initial and final state. Considering Heisenberg Hamiltonian of isotropic exchange interaction, the energy difference is $8JS^2$. Thus, the two-magnon energy-scale is a function of both the relevant exchange interaction and the magnitude of the spin. With the help of linear spin wave theory and the two-magnon Raman scattering, we can determine the exchange interactions of the systems.

**Figure 8.** Principle of two-magnon Raman scattering in the context of an antiferromagnetic spin chain.

Finally, we would like to point out that the polarizations of incoming and scattered photons play a major role to rightly determine the symmetry of the magnon modes. The effective two-magnon Raman scattering operator can be written as

$$\hat{o} = \sum_{i,j} \eta_{ij} (\hat{e}_i \cdot \vec{d}_{ij}) (\hat{e}_s \cdot \vec{d}_{ij}) \vec{S}_i \cdot \vec{S}_j \qquad \text{Eq. (7)}$$

, where $\eta_{ij}$ is the function of exchange interactions between spins on sites $i$ and $j$ [19]; $\hat{e}_i$ and $\hat{e}_s$ are the polarization vectors of the incident and scattered lights, respectively; $\vec{d}_{ij}$ is the connecting unit vector between the spin sites; $\vec{S}_i$ is the spin at the $i$-th site. Thus, the Raman scattering response strongly depends on the photon polarization directions with respect to the arrangement of magnetic ions in the crystal. If we now consider our antiferromagnetic spin chain, then finite two-magnon Raman scattering occurs when polarization of both incoming and scattered photons is parallel to the connecting vectors of the magnetic ions, i.e. parallel to the exchange coupling path. Hence, in this case, the symmetry of two-magnon Raman scattering is of $A_g$ symmetry.

## 6. Example of magnetic Raman scattering experiments
### 6.1. Two-magnon Raman scattering in antiferromagnetic Sr₂IrO₄

Sr$_2$IrO$_4$ is a Mott insulator driven by strong spin-orbit coupling and moderate electron-electron correlations. Here, Ir$^{4+}$ ions have the electron configurations of $5d^5$. Strong spin-orbit coupling associated with the $5d$-orbitals leads to total angular momentum of $j_{eff} = 1/2$. Such pseudospins $j_{eff} = 1/2$ form a two-dimensional antiferromagnetic order analogous to the well-known antiferromagnetic state of La$_2$CuO$_4$ below the Néel temperature of $T_N = 240$ K, as shown in Figure 9. In the following, we will review a Raman scattering study of two-magnons in this compound.

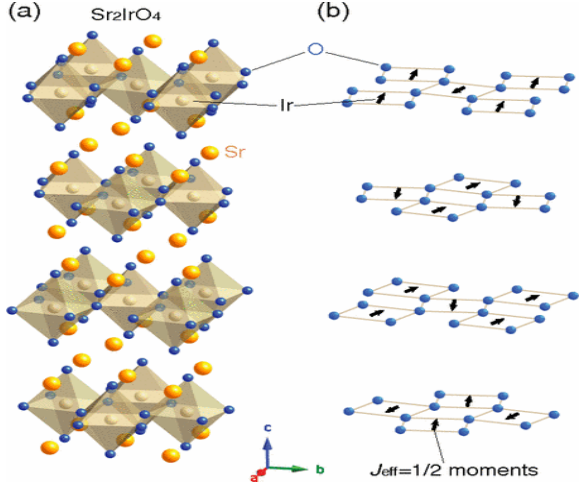

**Figure 9.** (a) Single crystal unit cell of Sr$_2$IrO$_4$. Networks of IrO$_6$ octahedra form layered structure. (b) 2D Heisenberg antiferromagnetic order in the layered structure. The figure is adapted from [20].

Figure 10(a) shows polarized Raman spectra at several temperatures from a c-axis aligned Sr$_2$IrO$_4$ crystal. The colored solid-lines represent the spectra taken in crossed polarization configuration, -c(ab)c in Porto's notation [propagation direction of incident light (polarization of incident light and polarization of scattered light) propagation direction of the scattered light]. Whereas the black solid-line represents the spectrum taken at 10 K in -c(aa)c mode. The former probes the Raman modes of B$_{2g}$ symmetry, and the later probes a combination of A$_{1g}$ and B$_{1g}$ symmetry considering the tetragonal symmetry of the crystal, i.e.

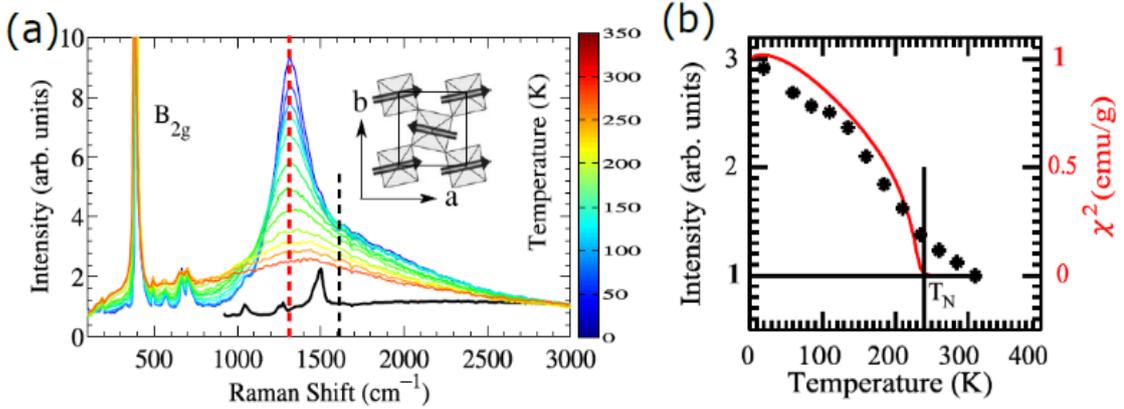

**Figure 10.** (a) Raman scattering response of $B_{2g}$ symmetry from Sr$_2$IrO$_4$ single crystals at several temperatures. The spectrum in black was recoded at 10 K and has $A_{1g} + B_{1g}$ symmetry. (b) Integrated two-magnon intensity as a function of temperature and the DC magnetic susceptibility ($\chi$) as a function of temperature. The figure is adapted from [35].

$D_{4h}$ point group. The broad mode around 1300 cm$^{-1}$ in B$_{2g}$ symmetry gradually evolves towards low temperatures. In particular, the intensity evolution of the mode clearly follows the temperature dependence of magnetic susceptibility, as shown in Figure 10(b). Further the mode persists even above the transition temperature. Hence, we can associate this mode with the two-magnon Raman scattering in this antiferromagnetic compound. Notably, the mode is not present in A$_{1g}$+B$_{1g}$ symmetry. Therefore, the magnon mode has only B$_{2g}$ symmetry. The other modes below 1000 cm$^{-1}$ are the phonon modes.

At first, we discuss a method on how easily we can estimate the energy-scale of two-magnon from the underlying exchange interactions. Earlier, resonant inelastic x-ray scattering study at the Ir $L_3$-edge obtained spin wave dispersion of the system [20]. Linear spin wave theory could correctly fit such spin wave dispersion with the following parameters for the exchange interactions. Nearest neighbor exchange coupling $J = 60$ meV, next-nearest neighbor interaction $J' = -20$ meV, and the third nearest neighbour coupling $J'' = 15$ meV. As

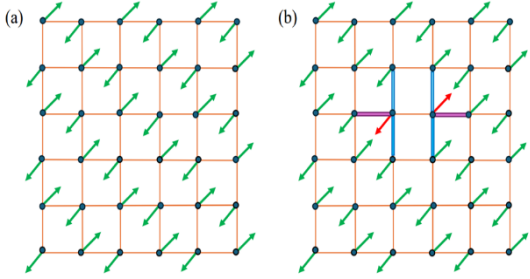

mentioned earlier, we can view two-magnon Raman scattering as the local spin exchanges between neighboring atomic sites. Such spin flips cost energy. Figure 11(a) shows 2D Heisenberg antiferromagnetic order. Creation of two-magnons in this ordered state leads to two local spin flips, as indicated by the red arrows in Figure 11(b). This frustrates the surrounding exchange interactions. Here, we have only highlighted the affected bonds that are nearest neighbor. The same applies to the next-nearest neighbor and the third nearest neighbor. According to Heisenberg Hamiltonian, $H =$

**Figure 11.** (a) 2D Heisenberg antiferromagnetic order. (b) Two local spin flips associated with two-magnons affect the surrounding exchange coupling bonds. Here, the affected nearest neighbor bonds are highlighted.

$\sum_{i,j} J_{ij} \vec{S}_i \cdot \vec{S}_j$, the total energy cost amounts to $3J - 4J' - 4J''$. In Since $J'$ and $J''$ are similar in magnitude, but opposite in sign, they cancel each other. Thus, effectively the two-magnon mode should appear at $3J$. However, 2D Heisenberg magnets of $S = 1/2$ with only the nearest neighbor exchange interaction leads to quantum fluctuations that reduce the two-magnon energy to $2.7J$ [21].Thus, the two-magnon mode should appear at 1306 cm$^{-1}$ which agrees well with the experimental results.

Having discussed the energy-scale, we now focus on the linewidth and line shape of the two-magnon mode. Overall, the linewidth of the mode broadens with increasing temperature. This is expected as the population of magnons (bosons) increases with increasing temperatures. Thereby, larger magnon-magnon scattering at elevated temperatures reduces lifetime of magnons and increases linewidth of the magnons. Another decay mechanism for the magnons will be via phonons. Increasing phonon density with increasing temperature also broadens the linewidth of the magnons. Further, the two-magnon mode is not symmetric Lorentzian-like. Rather, the mode has a long tail towards higher energy. This is associated with the fluctuations in the amplitude of the magnetization order parameter. This is known as the Higgs mode [21]. Note that two-magnon Raman scattering involves several magnons of opposite momenta. Hence, we cannot particularly discuss the linewidth of a single magnon.

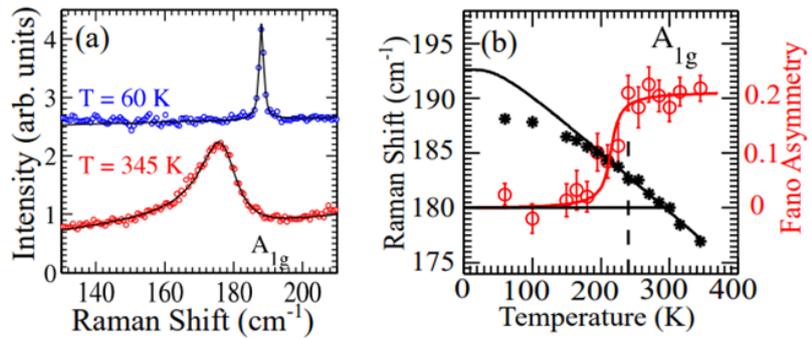

**Figure 12.** (a) $A_{1g}$ phonon mode below and above the Néel temperature. (b) Raman shift and Fano asymmetry of the $A_{1g}$ mode as a function of temperature. The figure is adapted from [35].

After discussing the two-magnon mode, we want to discuss about the temperature-dependent phonon lineshape of one $A_{1g}$ phonon mode displayed in Figure 12(a). The phonon line is narrow and Lorentzian-like at the low temperatures. However, at room temperature, such a phonon mode gets broader. Moreover, it can no longer be fitted with a Lorentzian profile. Specifically, the mode shows a significant asymmetry towards the low-energy side (lower Raman shift). Such a phonon lineshape closely resembles the Fano profile of phonons that couple to electronic continuum scattering. The Fano profile of phonons arises from the interference of discrete vibrational excitations with simultaneous continuum scattering. Such continuum scattering may originate from different sources. Here, the continuum scattering, as confirmed by the existence of Fano lineshape of the phonon does not occur from electronic continuum scattering as $Sr_2IrO_4$ is an insulator. The next possible source of continuum scattering might be related to pseudospin excitations. However, this must be rather spin fluctuations in the paramagnetic state. Figure 12(b) further shows temperature dependence of the peak position of the mode. At higher temperatures, the energy-scale follows simple symmetric anharmonic phonon decay model. Near Néel temperature, it deviates from this model. A prominent change occurs in the Fano asymmetry as a function of temperature. The asymmetric lineshape sharply diminishes below Néel temperature. This indicates that expected spin fluctuations form a long-range order in the antiferromagnetic state. Thus, the continuum Raman scattering evolved into sharp two-magnon modes.

In the context of phonon line shape, we would like to add that a Fano profile for Raman phonon modes is a typical signature of microscopic interactions between phonons and other degrees of freedom, such as spins and electrons. Additionally, in various ordered magnets, although the phonon modes do not show a Fano profile, their line shape as a function of temperature does not follow the typical symmetric anharmonic phonon decay (where the linewidth should decrease as a function of temperature when only phonon-phonon interaction is prevalent). Specifically, around the magnetic transition temperature, the phonon linewidth shows an anomaly in the phonon linewidth versus temperature curve. This is also a fingerprint of spin-phonon coupling. Details related to this can be found in references [22], [23].

## 6.2. Magnetic Raman scattering from $La_2CuO_4$

t would be incomplete without the discussion of magnetic Raman scattering from the parent cuprate compound La$_2$CuO$_4$. In fact, the theory of Raman scattering from correlated systems was developed based on the experimental results of this compound [3]. La$_2$CuO$_4$ is an antiferromagnet where the predominant exchange interactions occur between the Cu ions within the CuO$_2$ planes, with a coupling strength of ~135 meV. However, the coupling between the CuO$_2$ planes along the $c$-axis is considerably weak [24]. Consequently, the Néel temperature associated with its 3D antiferromagnetic order is not very high, precisely 325 K [25]. Figure 13(a) shows 3D antiferromagnetic order of the copper $s = 1/2$ spins. The magnetic structure is like the one of Sr$_2$IrO$_4$. In 1988, Lyons et al. obtained a broad two-magnon Raman scattering of $B_{1g}$ symmetry from La$_2$CuO$_4$ single crystals around 3000 cm$^{-1}$, as shown in Figure 13(b). As discussed earlier for a 2D Heisenberg $s = 1/2$ system, the two-magnon mode appears at $2.7J$. Thus, $J$ amounts to ~1100 cm$^{-1}$ (136 meV) which agrees with the neutron scattering results. Further, the two-magnon modes for both excitations lasers of 514.5 and 488.0 nm appear at the same position, as expected for inelastic scattering. However, their difference in shape is due to laser-dependent weak resonance features, as the Raman scattering involves intermediate transitions to virtual states.

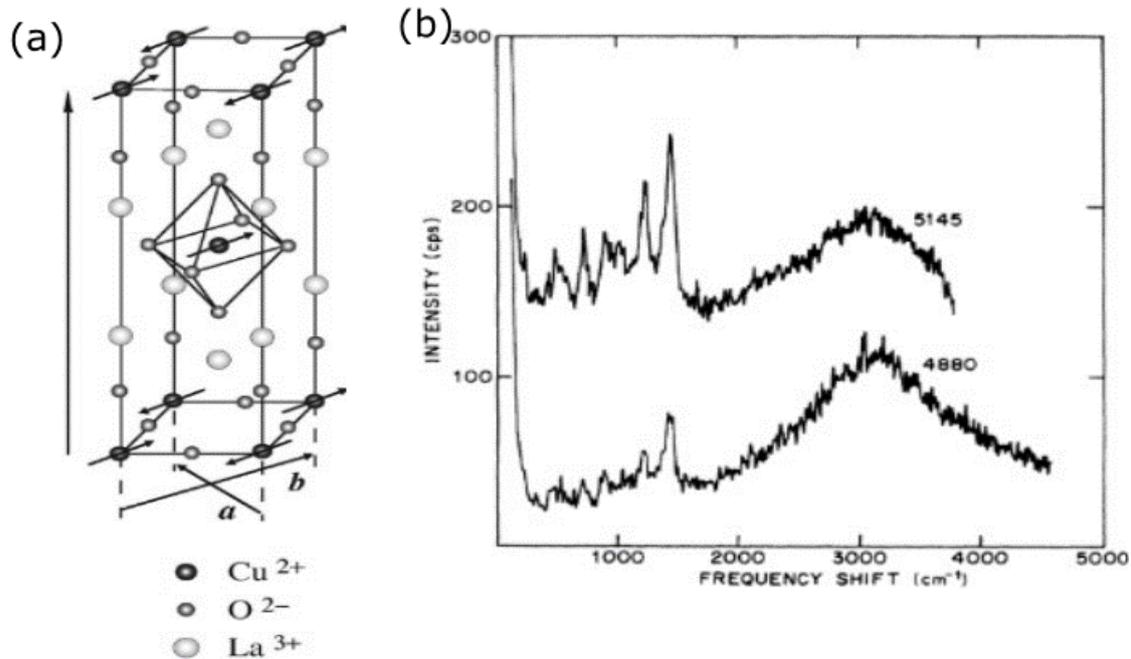

**Figure 13.** (a) 3D antiferromagnetic order of Cu $s = 1/2$ spins in a La$_2$CuO$_4$ unit cell. The figure is taken from ref. [36]. (b) Raman spectra measured in $yy$-geometry. Two excitation lasers of 514.5 and 488.0 nm wavelengths were used. Figure is adapted from [37].

### 6.3. One-magnon scattering from a monolayer CrI$_3$

2D materials play a major role in the discovery of exotic quantum ground states that occur in reduced dimensions. Simple available exfoliation techniques can easily achieve a few monolayers to single monolayer materials. Thus, these materials serve as natural thin films without any thin-film-growth related defects. In this section, we discuss Raman scattering experiments from a monolayer CrI₃ with emphasis on chirality of the magnon mode.

The van der Wall (vdW) compound CrI₃ is an Ising ferromagnet with all the electrons spins oriented in perpendicular to its vdW planes. The bulk compound has the Curie transition temperature ($T_c$) of 61 K. Remarkably, a single monolayer of CrI₃ is still a ferromagnet with reduced $T_c = 45$ K [26].The bilayer compound develops an antiferromagnetic order which involves an antiferromagnetic coupling of the two ferromagnetic vdW planes. The system returns to the bulk ferromagnetic state with the critical thickness of three vdW layers.

In monolayer CrI₃, Raman scattering experiments with the laser excitation of 632.8 nm obtained one-magnon mode from the acoustic magnon branch [27] for which magnon energy increases with its wavevectors, as shown in Figure 3. Since Raman scattering probes one magnons at the Brillouin zone center, the actual Raman response gets overwhelmed with the bright elastic Rayleigh scattering that puts an obstacle to probe very low-energy Raman response. With the help of optical notch filters low-frequency Raman magnetic mode was resolved under external magnetic fields. The role of external magnetic field is to raise the one-magnon energy and keep it away from the bright elastic Rayleigh line. Figure 14 shows Raman scattering response from a single monolayer at 15 K under magnetic field. The scattering response was recorded under the restriction of circular polarization of the incident and outgoing photons. At zero magnetic field, we did not observe any low energy magnon mode, as this is expected to get merged with the Rayleigh elastic line. When the magnetic field of -4T is applied the Stokes magnon part gives rise to a decent intensity in a crossed polarization configuration that involves right circular incident light and left circular scattered light. When the incident polarization is left-circularly polarized, the crossed-polarization configuration resolves the anti-Stokes part of the magnon. The opposite is true when we flipped the direction of the applied magnetic field. We will revisit this interesting polarization dependence a little later. First, we will plot magnon energy as a function of applied magnetic field to extrapolate and determine the magnon energy at zero magnetic field. There is a linear relationship between the applied magnetic field and the energy of the magnon mode (see Eq. (3) for the corresponding Hamiltonian). This allows use to determine the magnon energy or the gap in the spin-wave dispersion at the Brillouin zone center. We obtained the magnon gap at the zone center is 2.4 cm$^{-1}$ ($\sim 0.3$ meV). This magnon gap is quite large for ferromagnets that arise from the out-of-plane magnetic anisotropy due to vdW structure.

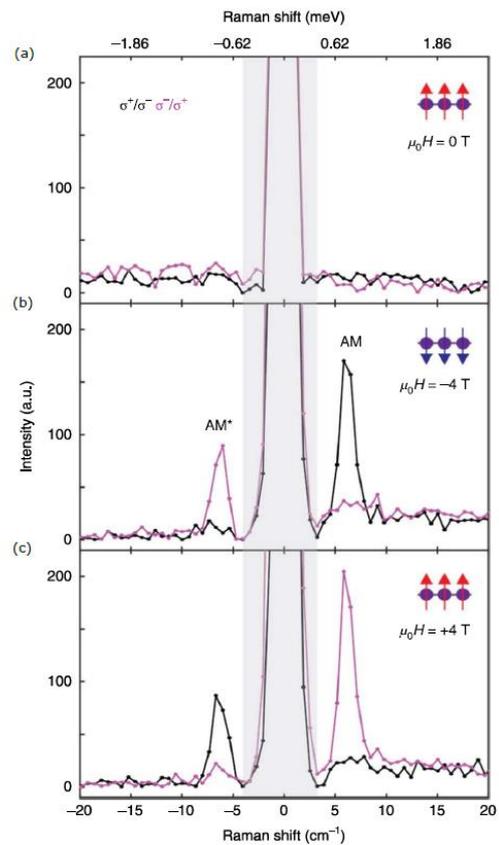

**Figure 14.** Stokes and anti-Stokes magnon modes from monolayer CrI₃ in crossed-circular polarization configurations. $\sigma^+$ and $\sigma^-$ stand for right and left circular polarized light, respectively. The figure is taken from [27].

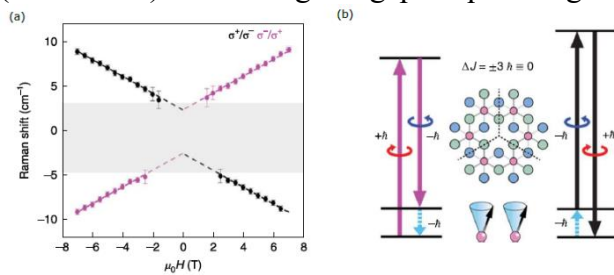

Now, we come to the polarization selection rules for observing the Stokes and anti-Stokes magnons. In monolayer CrI$_3$, the crossed-circular-polarization selection rules arise from the conservation of angular momentums of photons ($J_{ph}$) and magnons ($J_m$) within a honeycomb lattice. When the magnetization is directed up, the photon angular momentum changes by $\Delta J_{ph} = \pm 2\hbar$ as the helicity of the light reversed upon scattering, as shown in Figure 15(b) [27]. The positive sign holds for anti-

**Figure 15.** (a) Central energy of the Raman magnon modes as a function of applied magnetic field. The legends indicate the two crossed-circular-polarization configurations. (b) The angular momentum selection rules for one-magnon scattering in CrI$_3$ monolayer. The left (right) energy diagram is for Stokes (anti-Stokes) scattering. Total angular momentum changes of $\pm 3\hbar$ is equivalent to 0 in the honeycomb lattice. In the monolayer, Cr$^{3+}$ is pink and I$^-$ ions are green and blue. The figure is adapted from [27].

Stokes and the negative sign applies to Stokes scattering. Upon light scattering, the magnon angular momentum changes by $\Delta J_m = \pm \hbar$. Thus, the total change of angular momentum is $\Delta J = \Delta J_{ph} + \Delta J_m = \pm 3\hbar$. Any systems with continuous rotational symmetry would forbid such a change in angular momentum. However, three-fold rotation symmetry in the honeycomb CrI$_3$ permits angular momentum conservation up to $|3\hbar|$.

### 6.4. Magnetic Raman scattering from fractionalized excitations

In the preceding section, we discussed long-range antiferromagnets and ferromagnets, where magnetic Raman scattering produces sharp modes in the magnetic phases. In this section, we will discuss magnetic Raman scattering from systems with large quantum fluctuations, which prevent long-range order even at low temperatures. The characteristic feature of Raman scattering from such systems is the broad background in the Raman spectra, known as continuum scattering. Here, we discuss Raman scattering results from the two Kitaev-type quantum spin liquid candidate materials α-RuCl$_3$ and β-Li$_2$IrO$_3$. However, the results apply to any quantum spin liquid candidates.

#### 6.4.1. Magnetic Raman scattering from α-RuCl$_3$

It has been predicted that strongly spin-orbit coupling driven Mott insulators with the magnetic ions arranged on honeycomb lattice may harbour 2D QSL ground state. A prominent candidate of this class is α-RuCl$_3$ [28]. This crystallizes in a layered structure with plane of edge-sharing RuCl$_6$ octahedra arranged on a honeycomb lattice, as shown in Figure 16. Thus, the magnetic Hamiltonian of the system should be close to the Heisenberg-Kitaev model. Previous magnetic susceptibility and specific heat measurements obtained the signatures of two magnetic phase transitions at 8 K and 14 K [29], [30].

Figure 17(a) shows polarized Raman geometry for the measurements of α-RuCl₃ crystals. Figure 17(b) shows Raman spectra at 5 K in $xx$ and $xy$ polarization configurations[31]. The spectra are predominantly composed of sharp phonon modes, indicating the good crystallinity of the samples. Further, near low-energy below 20 meV, a broad mode exists, as highlighted in Figure 17(c). In case of insulating samples like α-RuCl₃ with

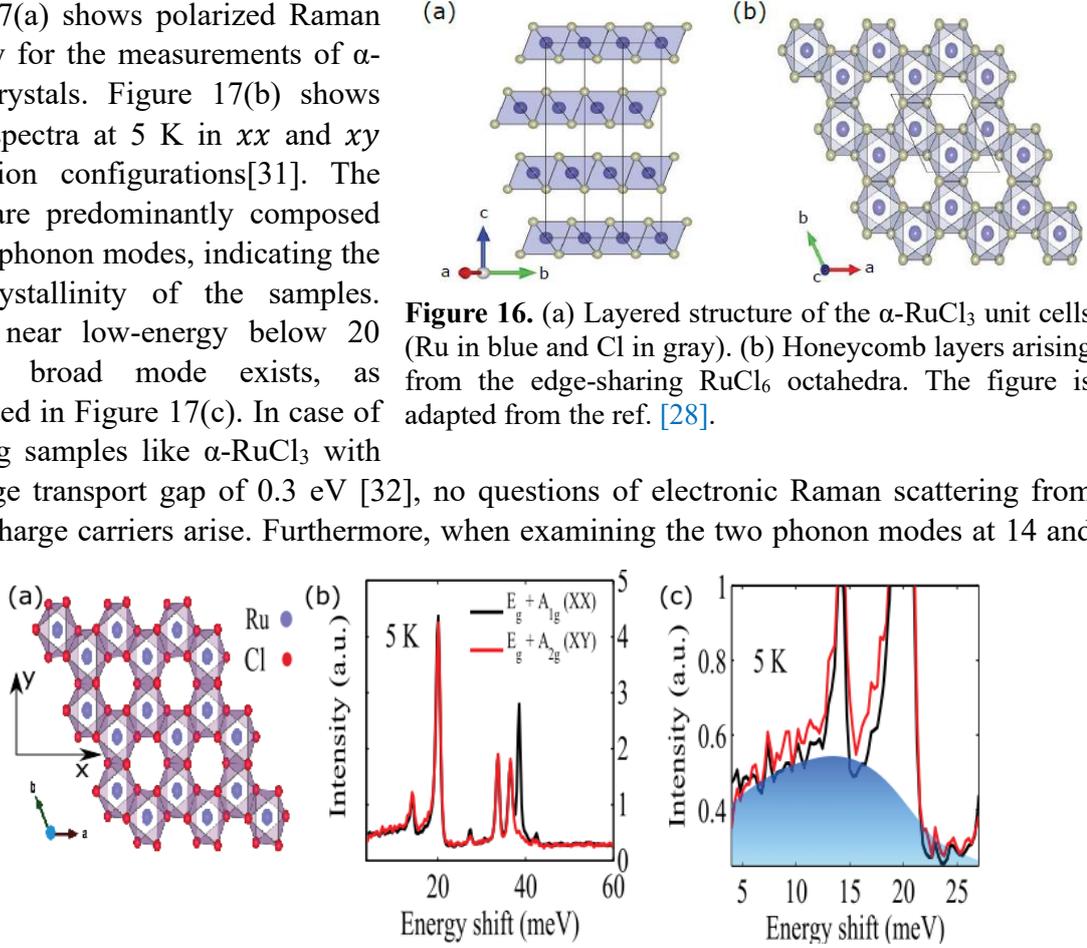

**Figure 16.** (a) Layered structure of the α-RuCl₃ unit cells (Ru in blue and Cl in gray). (b) Honeycomb layers arising from the edge-sharing RuCl₆ octahedra. The figure is adapted from the ref. [28].

the charge transport gap of 0.3 eV [32], no questions of electronic Raman scattering from mobile charge carriers arise. Furthermore, when examining the two phonon modes at 14 and

**Figure 17.** (a) Raman scattering geometry is given in $xy$-coordinate with respect to a vdW layer. (b) Raman spectra obtained in $xx$ and $xy$ polarization configurations at 5 K. (c) Finite Raman scattering continua up to 25 meV is highlighted. The figure is adapted from ref. [31].

20 meV, we observe a strongly asymmetric linewidth for the phonons. This asymmetry resembles Fano asymmetry, indicating that the phonons are coupled to a non-resonant scattering continuum. In this case, the authors argue, based on the preceding two arguments, that the scattering continuum, which persists at least up to 25 meV, arises from magnetic excitations in the system. In long-range magnets, magnons give rise to sharp modes, unlike those reported here. Furthermore, the authors conducted temperature-dependent Raman scattering experiments to identify the source of this broad magnetic scattering continuum.

Figure 18 (a) shows low-energy Raman spectral intensity at several temperatures in $xy$ polarization geometry. In conventional phonon or electronic Raman scattering, the spectral intensity should continuously decrease with decreasing temperature. However, we observed that the spectral intensity does not decrease with decreasing temperature below 100 K.

To identify the precise temperature dependence, spectral weight was calculated following the formula $SW = \int_{2.5\,\text{meV}}^{12.5\,\text{meV}} I(\omega) d\omega$, where $I(\omega)$ is the energy-dependent $(\omega)$ intensity. As shown in Figure 18(b), $SW$ decreases with decreasing temperatures down to 100 K followed by a slight increase below 100 K. For conventional phonon or electronic Raman scattering, $SW$ must follow the thermal Bose factor $n(\omega) = \frac{1}{e^{\hbar\omega/k_B T}-1}$. In particular, $SW$ should follow $[n(\omega) + 1]$ or $[n(\omega) + 1]^2$ for one- or two-particle scattering, respectively. However, it is not the case for α-RuCl₃. The sizable $SW$ below 100 K indicates that the scattering evolves from quasiparticles that do not follow Einstein-Bose statistics. The authors argue that the scattering continua in this system arise from fractionalized excitations. Note that, to prove fractionalized excitations in Raman spectra, we indeed need to measure down to low temperatures, as low as possible. For this, we need to measure Raman response for long hours with minimum laser power. In addition, we should use a phonon mode as Raman thermometry. Basically, we need to measure both Stokes and anti-Stokes for the selected phonon mode. Using their intensity ratio that follows Boltzmann distribution at a temperature, we can determine actual temperature upon laser illumination for Raman scattering. This is also discussed in section 4. A crucial advantage of Raman scattering over inelastic neutron scattering (INS) is that Raman scattering can obtain sizable Raman response from thin films down to a few nanometers thickness, or tiny single crystals. However, INS requires much bigger samples. If the sample has a strong neutron absorbing element like Ir, we cannot perform INS. In Raman scattering, we do not have such a problem. The further advantage of Raman scattering is its unparallel energy resolution up to 0.06 meV, whereas it is >=1 meV in INS. Moreover, Raman scattering is a home-laboratory based technique. However, the disadvantage is that INS can measure as a function of crystal momentum, however Raman scattering yields average response from the selected part of Brillouin zone.

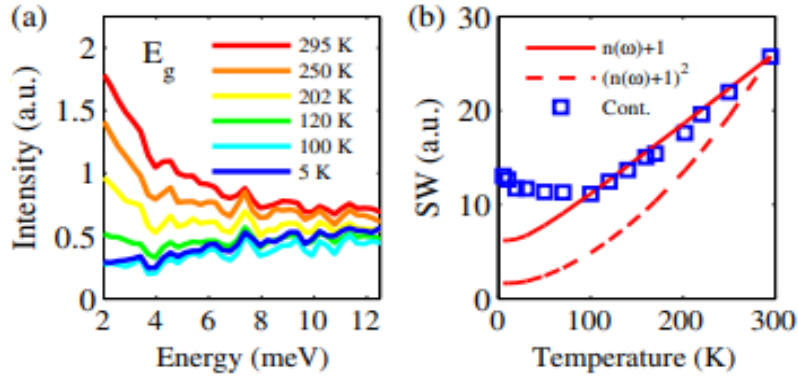

**Figure 18.** (a) The low energy Raman spectra in $xy$ −polarization configurations at various temperatures. (b) The integrated spectral weight ($SW$) at low energy vs. temperature (shown with symbols). The lines illustrate the expected spectral weight for both single and two-particle scattering from bosons. The figure is adapted from [31].

### 6.4.2. Magnetic Raman scattering from β-Li₂IrO₃

β-Li₂IrO₃ is a three-dimensional honeycomb lattice, unlike α-RuCl₃ where honeycomb network of the magnetic ions lies in the $ab$-plane. As shown in Figure 19(a) alternating exchange interactions form twisted zigzag chains. These zigzag chains are further interconnected through another exchange interaction. Figure 19 (b) shows Raman scattering response of single crystalline samples at several temperatures in a parallel polarization configuration where the polarizations of incoming and outgoing photons are both parallel to the crystallographic $c$-axis. First, the Raman spectra show finite continuum scattering that develops into a quasi-elastic peak (spectral weight around zero Raman shift) at higher temperatures. Such finite scattering continua arise from dynamical spin fluctuations in paramagnetic state with short-range spin correlations. To precisely determine the temperature-dependent evolution of the scattering continuum, the integrated intensity $(I(\omega))$ over the energy range from 25 to 51 meV was plotted as a function of temperature in Figure 19 (c). $I(\omega)$ was fitted reasonably well with the scattering from Bosons and two fermions. Here, the boson scattering is associated with the phonon scattering background and/or the magnons. Whereas, the two-fermion scattering is associated with the creation and annihilation of pairs of fermions. The bosonic background is defined by $[n(\omega) + 1]$ and the two-fermion scattering is modelled following $[1 - f(\omega)]^2$ where $f(\omega) = 1/(1 + e^{\hbar\omega/k_B T})$ [11]. Note that in absence of any fermionic scattering, the boson scattering would be enough to fit this integrated intensity. In the following we will discuss the nature of these fermions by analysing phonon lineshapes.

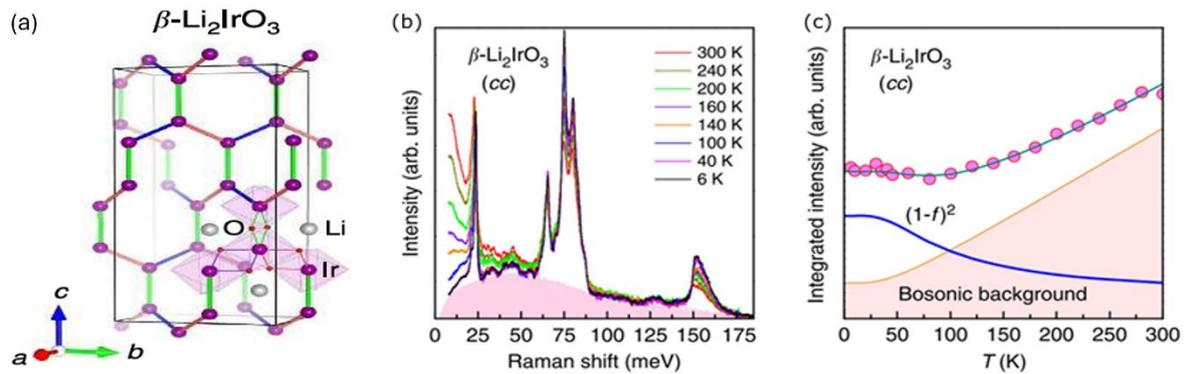

**Figure 19.** (a) Single crystal unit cell of β-Li2IrO3, emphasizing the exchange interaction network and highlighting the Kitaev network. (b) Raman spectra in a parallel polarization configuration at several temperatures. The shaded area represents the considering bosonic background and two-fermion scattering. The figure is adapted from ref. [38].

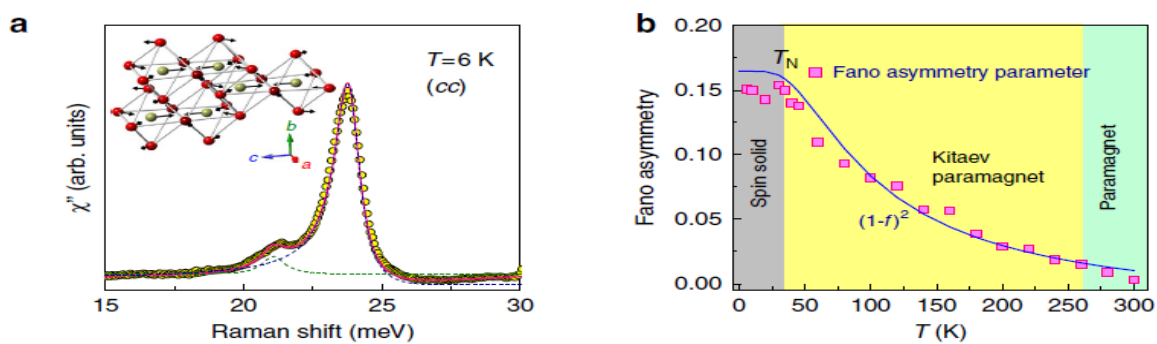

**Figure 20.** (a) The asymmetric phonon line shapes along with the Fano profile fitting. (b) Fano asymmetry parameter as a function of temperature (symbols). The solid line through the data indicates dominant two-fermion scattering. The figure is adapted from ref.[38].

Further, the existence of strongly asymmetric phonon at ~24 meV indicates the Fano-like feature arising from the coupling of the phonon mode to the continuum excitations [33]. To fit this asymmetric phonon, first imaginary part of the Raman scattering response was calculated using $\chi''(\omega) = I(\omega)/[1 + n(\omega)]$. Then the phonon mode is fitted with a Fano profile $I(\omega) = I_0 (q + \varepsilon)^2/(1 + \varepsilon^2)$, as shown in Figure 20(a). The reduced energy term $\varepsilon$ is defined as $\varepsilon = (\omega - \omega_0)/\Gamma$, where $\omega_0$ is the intrinsic phonon frequency without coupling, $\Gamma$ is the linewidth, and $q$ is the Fano asymmetric parameter. The asymmetric lineshape is quantified by the parameter $1/|q|$, i.e. the parameter increases with the asymmetry in the phonon lineshape. Figure 20 (b) shows the Fano asymmetry parameter as a function of temperature. This rises with decreasing temperature and becomes a constant below the transition temperature for the long-range antiferromagnetic order. Quite surprisingly, such a temperature-dependence closely matches with the two-fermion scattering function $[1 - f(\omega)]^2$. The two-fermion scattering function describes the temperature-evolution of the integrated intensity of continuum Raman scattering as well. A reasonable interpretation of such resemblance is that the spins in Kitaev honeycomb systems thermally fractionalize into itinerant Majorana fermions [11]. Therefore, the continuum Raman scattering arises from the fractionalized excitations that are coupled to the phonon modes. Consequently, Raman scattering can serve as a suitable probe for fractionalized excitations, capturing not only continuum scattering from them but also the existence of conventional bright phonon modes through their asymmetric lineshape.

## 7. Summary and perspectives

We have demonstrated the potential of polarized Raman scattering through several experimental results across a wide range of quantum materials. Our focus was on two primary classes of materials: (1) magnetic materials with long-range ferromagnetic and antiferromagnetic orders, and (2) quantum spin liquids characterized by strong quantum spin fluctuations that prevent long-range magnetic orders even at low temperatures. The signatures of these quantum magnetic states are low-energy excitations, specifically magnons in ferromagnets and antiferromagnets, and fractionalized excitations in quantum spin liquids.

Through various examples of Raman scattering studies on relevant quantum materials, we established that Raman scattering is an excellent tool for probing both types of quasiparticles. Magnons in Raman scattering exhibit two distinct types of signatures: one-magnon scattering, which is relatively sharp and appears at very low energy, corresponding mainly to the magnon gap at the Brillouin zone center; and two-magnon scattering, which is a broader mode that typically appears at higher energies, representing the magnon density of states.

Our studies further showed that the energy of the magnon modes can be used to determine exchange interactions and magneto crystalline anisotropy gaps at the zone center. We provided detailed explanations of the mechanisms behind one-magnon and two-magnon scattering, highlighting the necessity of spin-orbit coupling for observing one-magnon scattering. One-magnon scattering is common in ferromagnets, whereas two-magnon scattering is typically observed in antiferromagnets.

We also emphasized the importance of the polarization of incoming and scattered photons in Raman scattering. By carefully selecting these polarizations, we can obtain Raman scattering responses of specific symmetries, enabling determination of polarization-resolved quasiparticle dynamics. Specifically, we discussed two-magnon scattering from single crystals of antiferromagnetic $Sr_2IrO_4$ and $La_2CuO_4$, which are structurally and magnetically analogous.

Additionally, we presented one-magnon scattering results from the der Waals ferromagnetic CrI3 monolayers. One notable advantage of Raman scattering is its ability to simultaneously probe both magnons and phonons in materials. The lineshape analysis of the phonons can reveal spin-phonon coupling, for instance.

In our review, we also highlighted fractionalized excitations in Kitaev spin liquid candidates. The signature of these quasiparticles in Raman scattering is the Raman spectral background, known as Raman continuum scattering. The origin of these continua from fractionalized excitations was confirmed through temperature-dependent Raman scattering responses, showing that the Raman spectral weight of the continuum background follows Fermi-Dirac statistics for the creation or annihilation of fermion pairs. Remarkably, phonons can interfere with these fractionalized excitations too, resulting in Fano asymmetry in the phonon lineshape.

Overall, our review demonstrated that Raman scattering is an efficient technique capable of simultaneously probing various quasiparticles. The simultaneous occurrence of different quasiparticles allows us to detect couplings among them, such as spin-phonon coupling. Unlike inelastic neutron or x-ray scattering, which require large-scale facilities, Raman scattering is a laboratory-based technique utilizing inexpensive monochromatic visible lasers. Additionally, Raman scattering does not require large single crystals; micron-sized samples are sufficient to record Raman responses. Using confocal geometry, we can even record responses from samples only a few nanometers thick.

Our review aims to inspire young experimental scientists by showcasing the potential of Raman scattering in understanding quantum materials. We hope this motivates them to explore this technique in their research endeavors.

## Acknowledgement

K.S. acknowledges the financial support from the DST INSPIRE Faculty Fellowship, Core Research Grant from the SERB, and YSRP Grant from the BRNS, India.